\documentclass{ws-procs9x6-cpt25}
\usepackage{xcolor}
\usepackage{amsmath,amssymb}

\begin{document}

\newcommand{\refeq}[1]{(\ref{#1})}
\def\etal {{\it et al.}}
%any other macros go here 

\title{Bumblebee Gravity - Lessons from Perturbation Theory\footnote{The contents of these proceedings are based on Ref.\ \refcite{bumblebeeIBS} and was presented at CPT'25 in an earlier, preliminary form.}}

\author{N.A.\ Nilsson,\footnote{nilsson@ibs.re.kr}$^{1,2}$}

\address{$^1$Cosmology, Gravity and Astroparticle Physics Group, Center for Theoretical Physics of the Universe, Institute for Basic Science, \\ Daejeon 34126, Republic of Korea}
\address{\vspace{-3mm}$^2$LTE, Observatoire de Paris, Université PSL, CNRS, LNE, Sorbonne Universit\'e, \\ 61 avenue de l’Observatoire, 75s014 Paris, France}

\begin{abstract}
These proceedings summarise some recent efforts in understanding a class of vector-tensor theories known as {\it bumblebee} models, which spontaneously break local Lorentz and diffeomorphism invariance. Using cosmological perturbation theory on an FLRW background, we find that for non-minimal coupling to gravity, the theory contains a ghost mode unless  degeneracy conditions are imposed, after which the model becomes a subset of generalised Proca theory, and the potential is then completely fixed by the background equations. We find a constraint on the bumblebee field from the speed of tensor modes on the order of $10^{-15}$. We go further to show that scalar perturbations do not propagate at the linear level, indicating that the theory is pathological around dynamical cosmological backgrounds, a result which is independent of the form of the potential.
\end{abstract}

\bodymatter

\section{Introduction}\label{sec:intro}
The Standard Model of cosmology, with a hot Big Bang followed by an inflationary phase, eventually evolving to the dark-energy dominated universe we live in today, is the prevailing cosmological model. Using the two main ingredients, general relativity and the cosmological principle, this model accurately describes the evolution from inflationary scales, where quantum effects dominate, to the formation and evolution of large-scale structure. That being said, the underlying theory of gravity, general relativity, is not without problems; for example, the cosmological constant problem sports a discrepancy of $55$ orders of magnitude when comparing measurements and predictions from QFT\cite{CC}, and the Hubble parameter tension has reached particle-physics standards with a discrepancy of $> 5\sigma$.\cite{H0tension} These issues are deeply unsatisfactory as they keep us from achieving a truly elegant understanding and description of the Universe from primordial times until the present day. Whether from a fundamental misunderstanding of gravity at cosmological scales, hitherto hidden instrument systematics (such possibilities are discussed in the literature), or some other reason, there is now ample reason to study modifications of general relativity and/or one or more sectors of the Standard Model of particle physics.

The ultimate goal of physics can in some sense be said to be finding a ``theory of everything'', a single theory which gives accurate predictions at all scales, from the Big Bang to the present day and from cosmological scales to subatomic particles. Such a theory, sometimes called ``quantum gravity'' has been eluding scientists for over fifty years. There are many quantum-gravity candidate theories in the literature, for example string theory, loop quantum gravity, causal dynamical triangulation, Ho\v{r}ava-Lifshitz gravity, and more, with the latter three claiming to resolve the non-renormalizability problem of general relativity (and thus allowing for canonical quantisation) whereas string theory attempts to unify all fundamental interactions. A feature which appears in several proposals to quantum gravity is that of {\it broken spacetime symmetries}, which was highlighted by Kostelecky \& Samuel in Ref.\ \refcite{KostSam1} where it was shown that local Lorentz symmetry can be spontaneously broken in string field theory. As such, an EFT framework known as the Standard-Model Extension was developed\cite{SME1} to help search for minute departures from exact local Lorentz, {\it CPT}, and diffeomorphism symmetry\footnote{For all current constraints, see the Data Tables \cite{datatables}}, and whilst several hints of violation have been found (see for example Refs\ \refcite{violation1}, \refcite{violation2}), it is not currently enough to claim a detection.

A popular vector-tensor theory which was first written down in 1989 in Refs.\ \refcite{bb1}, \refcite{bb2} and later found to be a vector subset of the Standard-Model Extension is known as the {\it bumblebee model}, which incorporates spontaneous violation of local Lorentz symmetry (and therefore also diffeomorphism symmetry). This model, thanks to its relative simplicity, has been the subject of a significant amount of study in the last decades\footnote{A search around the time of writing reveals over 180 papers in the last 10 years.}, especially in the context of compact objects such as Schwarzchild-like solutions\cite{casana, xu, QBH}, rotating solutions\cite{ding}, and more. In cosmology, the literature is more scarce, with FLRW and AdS solutions at the background level\cite{carlos}, cosmological tests with CMB data\cite{CMB}, anisotropic cosmological solutions\cite{aniso} and more. We note the existence of a more general vector-tensor theory known as generalised Proca theory\cite{GP} which, although not strictly related to spacetime-symmetry breaking contains several bumblebee models as subsets, and the literature on cosmology with generalised Proca is larger than that for the bumblebee model\cite{GPcosmo1, GPcosmo2, GPcosmo3}. In these proceedings, we summarise some recent work on the stability of the bumblebee model, and we identify a consistency condition necessary for a healthy theory. These proceedings are based on the results in Ref.\ \refcite{bumblebeeIBS}. We use $(-,+++)$ signature and units where $c=\hbar=1$ and $G=1/8\pi M_{\rm Pl}^2$.

\section{The model}
Our starting point is the most general bumblebee action with non-mininal coupling to gravity as
\begin{equation}\label{eq:action}
\begin{aligned}
		\mathcal{L}_B \sim &\frac{M_{\rm Pl}^{2}}{2} R - \frac{\gamma}{4}B_{\mu \nu}B^{\mu \nu} - V(B^2) + \eta B^2 \nabla_{\mu}B^{\mu} 
		\\
		&\quad\quad\quad\quad+ \xi B^\mu B^\nu R_{\mu\nu} +\sigma B^2 R 
		+ \varsigma (\nabla_{\mu}B^{\mu} )^{2} 
		+ \upsilon 
		\nabla_\mu B^\mu R,
\end{aligned}
\end{equation}
where we note that in contrast to most of the bumblebee literature, the action is written such that all couplings $\{\gamma, \eta, \xi, \sigma, \varsigma, \upsilon\}$ are dimensionless. Further, $M_{\rm Pl}$ is the Planck mass and $V_B$ is the potential for the bumblebee field $B^\mu$ which spontaneously breaks the diffeomorphism (and hence local Lorentz) symmetry; moreover, since all operators are marginal, the only scale is introduced through the potential. Other terms can be considered in Eq.~(\ref{eq:action}), but not all are independent and can be rewritten using e.g.
$$
\int d^4x \sqrt{-g}R_{\mu\nu}B^\mu B^\nu = \int d^4x \sqrt{-g}\left[\left(\nabla_\mu B^\mu\right)^2-\nabla_\mu B_\nu \nabla^\nu B^\mu + \partial(\hdots)\right].
$$
By varying the Lagrangian (\ref{eq:action}) we obtain the Einstein equations; these are lengthy and we do not display them here (see Eq.~2.10 in Ref.~\refcite{bumblebeeIBS}). The equation of motion for the bumblebee field is
\begin{equation}\label{eq:BEq}
\begin{aligned}
    \frac{\gamma}{2}\Box B_\mu-&\frac{1}{2}(\gamma+2\varsigma)\nabla_\nu\nabla_\mu B^\nu+(\xi+\varsigma)B^\nu R_{\mu\nu}-\eta B^\nu\nabla_\mu B_\nu\\&\quad\quad
    +\sigma B_\mu R+\eta B_\mu\nabla_\nu B^\nu
		-\frac{\upsilon}{2} \nabla_{\mu} R -B_\mu V^\prime(B^2) = 0 \,,
\end{aligned}
\end{equation}

\section{Bumblebee gravity on a homogeneous and isotropic background}\label{sec:flrwbg}
On a flat Friedmann-Lemaitre-Robertson-Walker (FLRW) background with the metric in terms of cosmic time
\begin{equation}
    ds^2 = -dt \otimes dt +a^2(t)\delta_{ij}dx^i\otimes dx^j,
\end{equation}
where $a(t)$ is the cosmic scale factor, the Friedmann equations read
\begin{equation}\label{eq:BGeqs}
    \begin{aligned}
	 &3M_{\rm Pl}^2H^2-6\eta \bar{B}_0^3H+9(2\sigma-\varsigma)\bar{B}_0^2H^2-3(\xi+2\sigma)H(\bar{B}_0^2)^{\dot{}}-\varsigma\dot{\bar{B}}_0^2\\&\quad+6(\xi+2\sigma+\varsigma)\bar{B}_0^2\dot{H} +2\varsigma \bar{B}_0\ddot{\bar{B}}_0+6 \upsilon \Big[ \dot{\bar{B}}_{0}  \dot{H} -H \ddot{\bar{B}}_{0} - 2 H ^2 \dot{\bar{B}}_{0} \\&\quad\quad +\bar{B}_{0} \big(3  H ^3 -\ddot{H} -4 H  \dot{H} \big)\Big] -V(-\bar{B}_0^2)-2\bar{B}_0^2V^\prime(-\bar{B}_0^2)=0,\\
    &M_{\rm Pl}^2(3H^2+2\dot{H})-2(\xi+2\sigma+\varsigma)\bar{B}_0\ddot{\bar{B}}_0-(2\xi+4\sigma+\varsigma)\dot{\bar{B}}_0^2- 2 \eta  \bar{B}_{0}^{2} \dot{\bar{B}}_{0}\\&\quad-(2\xi+2\sigma+3\varsigma)(2\bar{B}_0^2\dot{H}
	+2 \upsilon \Big[-\tfrac{d^{3}}{dt^{3}} \bar{B}_{0} -5 H  \ddot{\bar{B}}_{0}-\dot{\bar{B}}_{0} 
	\big(5  \dot{H} +3 H ^2 \big)\\&\quad\quad\qquad\qquad+9 \bar{B}_{0}  H 
	\big( \dot{H} +H ^2\big)\Big] +2(\bar{B}_0^2)^{\dot{}}+ 3 \bar{B}_0^2H^2) -V(-\bar{B}_0^2) =0
    \end{aligned}
\end{equation}
and the bumblebee equation of motion reduces to a constraint of the form
\begin{equation}\label{eq:BeqBG}
\begin{aligned}
&\varsigma(\ddot{\bar{B}}_0+3H\dot{\bar{B}}_0) + 3(\xi+2\sigma+\varsigma)\bar{B}_0\dot{H}+3(\xi+4\sigma)\bar{B}_0H^2\\&\quad\qquad\qquad-3\eta \bar{B}_0^2H-\bar{B}_0V^\prime(-\bar{B}_0^2) -3 \upsilon \big(\ddot{H} +4 H  \dot{H} \big)=0 \,.
\end{aligned}
\end{equation}
In the above equations, we have chosen a timelike ansatz for the bumblebee field i.e.
$$
B_\mu \to \{\bar{B}_0(t), \vec{0}\},
$$
which is a natural choice since the background metric is isotropic. Note here that the case $\bar{B}(t)= \text{constant}$, which is the standard assumption from the point of view of spontaneous spacetime-symmetry breaking, is included in the below results as a subset. In flat space, we have $R_{\mu\nu}=0$ and therefore $V^\prime_B=0$, but in general we can write the potential as $V_B=V_B(B^2\pm b^2)$. 

\section{Linear cosmological perturbations of the bumblebee action -- existence of a ghost mode}\label{sec:cosmopert}
We study cosmological perturbations of the model (\ref{eq:action}) around exact dS, and we decompose the bumblebee field as 
\begin{equation}
    B_{\mu} \rightarrow \{\bar{B}_0+\epsilon\delta B_0, \,\, \epsilon\partial_i\delta B_s+\epsilon\delta B_i^{(T)}\},
\end{equation}
where we note the existence of two scalar modes and one divergenceless vector mode (not all of which are dynamical). We write the metric in ADM form as
\begin{equation}
    ds^2 = -N(t)^2dt^2 + a(t)^2\gamma_{ij}(dx^i+N^i dt)(dx^j+N^j dt),
\end{equation}
which is perturbed in spatially-flat gauge as
\begin{equation}
    \begin{aligned}
        N(t) = 1+\epsilon \alpha, \quad N^i = \frac{\epsilon}{a}\left(\mathcal{B}^i+\partial^i\beta\right), \quad
        \gamma_{ij} = \delta_{ij}+\epsilon h_{ij},
    \end{aligned}
\end{equation}
where we have two scalar, one divergenceless vector, and two symmetric and trace-free tensor modes. We now perturb the action to second order in linear perturbations of Scalars ($\alpha$, $\beta$, $\delta B_0$, $\delta B_s$), Vectors ($\mathcal{B}^i$, $(\delta B^\perp)^i$), and Tensors ($\gamma_{ij}$). Not all of these degrees of freedom are dynamical and propagating. Below, we show the dynamics of tensors and scalars.
\subsection{Tensor perturbations}
At second order in tensor perturbations, the action can be written as
\begin{equation}
    S^{(2)}_T = \frac{M_{\rm Pl}^2}{8}\int  d^3x\,dt \, a^3\,\mathcal{K}_T\Bigg[\dot{h}_{jk}\dot{h}^{jk} - \frac{c_T^2}{a^2}\partial_i h_{jk} \partial^i h^{jk}\Bigg],
\end{equation}
where the kinetic and gradient coefficients reads
\begin{equation}
\begin{aligned}
&\mathcal{K}_T\equiv  \Big[1-2(\xi+\sigma)\tilde{B}_0^2 - \frac{2 \upsilon}{M_{\rm pl}} (\dot{\tilde{B}}_0+3H \tilde{B}_0) \Big],  \\&c_T^2 \equiv \frac{1}{\mathcal{K}_T} \Big[1-2\sigma\tilde{B}_0^2- \frac{2 \upsilon}{M_{\rm pl}} (\dot{\tilde{B}}_0+3H \tilde{B}_0) \Big] \,,
\end{aligned}
\end{equation}
where we have defined the dimensionless quantity $\tilde{B}_0 \equiv \bar{B}_0/M_{\rm Pl}$. In the small-coupling limit, the speed of tensor modes can be expressed as $c_T^2\approx1+2\xi\tilde{B}_0^2$, $\xi,\sigma, \upsilon\ll1$.

\subsection{Scalar perturbations}
After going to Fourier space and integrating by parts, we straighforwardly find that $\beta$ is non-dynamical and we integrate it out. Due to the presence of the coupling $\upsilon$, the quadratic action contains $\ddot{\alpha}$ and $\delta\ddot{B}_0$, which we eliminate by the variable redefinition $\tilde{\alpha}\equiv \alpha-\delta B_0/\tilde{B}_0$ and the introduction $Q\equiv\dot{\tilde{\alpha}}$, after which we can write, to quadratic order in scalar perturbation
\begin{equation}\label{eq:s24}
\mathcal{L}^{(2)}_{\rm tot} \equiv \mathcal{L}^{(2)}_2 + \mathcal{L}^{(2)}_3 + \lambda \left(
Q - \dot{\tilde{\alpha}}
\right) 
= \mathcal{L}^{(2)}_2 + \mathcal{L}^{(2)}_3 + \lambda
Q + \dot{\lambda} \tilde{\alpha} + {\rm T.D.}
\,,
\end{equation}
where $\lambda$ is a Lagrange-multiplier field imposing $Q=\dot{\tilde{\alpha}}$. We can now integrate out $\tilde{\alpha}$ and write the action in terms of the variables $\{\dot{\lambda},\lambda,\dot{Q},Q\}$. We can now write the Lagrangian as
\begin{equation}
    \mathcal{L}^{(2)}=\frac{a^3}{2}\int d^3x \, dt\Big[\dot{\mathcal{V}}_4^\dagger\mathbf{K}_4\dot{\mathcal{V}}_4+k\dot{\mathcal{V}}_4^\dagger\mathbf{F}_4\mathcal{V}_4-\mathcal{V}_4^\dagger\mathbf{X}_4\mathcal{V}_4\Big],
\end{equation}
where $\mathbf{K}_4$ is the kinetic (Hessian) matrix, $\mathbf{F}_4$ is the friction matrix, and $\mathbf{X}_4$ contains the gradient and mass matrices. Also, we have defined the vector $\mathcal{V}_4 \equiv \left(\lambda, Q, \delta B_0, \delta B_s\right)$. The kinetic matrix $\mathbf{K}_4$ is lengthy but can be found to have $\text{max}[\text{rank}(\mathbf{K}_4)]=3$, showing that some cancellations take place, since the rank could in principle be four. We now need to reduce the rank to unity in order to obtain the correct number of propagating scalar modes. We find that the only way to obtain a matrix rank less than three is to impose $\upsilon\to0$ through which we find $\operatorname{max[rank}(\mathbf{K}_4)|_{\upsilon=0}] = 2$. As such, we find that $\upsilon$ needs to vanish, and we must take
\begin{equation}
    \boxed{\upsilon=0}.
\end{equation}
Imposing this into the quadratic action (\ref{eq:s24}), we find that $\lambda$ can be integrated out and we find
\begin{equation}
    \mathcal{L}^{(2)}_2 = \frac{1}{2}a^3\left(
\dot{\mathcal{V}}_3^\dagger\mathbf{K}_3\dot{\mathcal{V}}_3 
+\dot{\mathcal{V}}_3^\dagger\mathbf{N}_3\mathcal{V}_3 
-\mathcal{V}_3^\dagger\mathbf{X}_3\mathcal{V}_3
\right),
\end{equation}
where $\mathcal{V}_3^\dagger \equiv (\alpha,\delta B_0,\delta B_s)$, and where the kinetic matrix can be written
\begin{equation}
    \mathbf{K}_3 = \begin{pmatrix}
        K \bar{B}_0^2 & -K \bar{B}_0 & 0 \\
        -K \bar{B}_0 & K & 0 \\
        0 & 0 & \gamma\frac{k^2}{a^2}
    \end{pmatrix},
\end{equation}
where 
\begin{equation}\label{eq:def-K}
K \equiv -\frac{2}{\varsigma} (\xi + 2 \sigma )(\xi +2\sigma+2\varsigma) \,,
\end{equation}
and where $\operatorname{rank}(\mathbf{K}_3) = 2$. This shows that the theory now have two propagating scalar degrees of freedom in general. From Eq.~(\ref{eq:def-K}) we see that there are two non-trivial possibilities which reduce the rank to unity. First of all, $K$ is singular in $\varsigma$, but this is spurious since the action (\ref{eq:action}) shows no such singularity. We can therefore go back to the action and set $\varsigma=0$ and again compute the quadratic action for scalar perturbations. We find that we can now integrate out both $\beta$ and $\alpha$, after which we have a rank-2 kinetic matrix of the form $\mathbf{K}_2=\text{diag}(18 (\xi +2 \sigma )^2 \frac{{H}^2 \bar{B}_0^2}{D}, \gamma  \frac{k^2}{a^2})$, where $D$ is lengthy. Since the rank has not been reduced to unity we need to impose further constraints to obtain a healthy theory; from Eq.~(\ref{eq:def-K}) we have the two possibilities
\begin{align}\label{eq:DC1}
&\xi +2\sigma =0  \quad \text{and} \quad \xi +2\sigma+2\varsigma \neq 0 \,,
\\ \label{eq:DC2}
&\xi +2\sigma \neq 0 \quad \text{and} \quad \xi +2\sigma+2\varsigma = 0 \,.
\end{align}
Considering both possibilities, we find that the necessary and sufficient condition for a single propagating scalar mode is
\begin{align}\label{eq:degencond}
&\boxed{\xi +2\sigma =0  \quad \text{and} \quad \varsigma = 0 } \,.
\end{align}
Therefore, we conclude that the above condition is the only choice which leads to the correct number of degrees of freedom, and taking limits e.g $\{\sigma=0, \xi\neq0\}$ is not possible\footnote{Issues with this case was also noted in Ref.~\refcite{baileybb} at the background level.}. In other words,
\begin{quote}
    {\it The bumblebee model (\ref{eq:action}) has a ghost instability unless the degeneracy condition (\ref{eq:degencond}) is imposed.}
\end{quote}
In fact, these degeneracy conditions are already known in the context of generalised Proca theory\cite{GP}; indeed,
with the degeneracy conditions (\ref{eq:degencond}) imposed, the bumblebee model becomes a subset of generalised Proca with the identification
\begin{equation}
\label{eq:procamap}
\begin{aligned}
&G_2 = -\frac{1}{4}B_{\mu\nu}B^{\mu\nu}-V_B\left(-2X\right), \quad G_3=-2\eta {X}, \\ &G_4 = \frac{M_{\rm Pl}^2}{2} + \xi X \,;
\quad
X \equiv -\frac{1}{2} B_\mu B^\mu \,,
\end{aligned}
\end{equation}
and at this point, all results obtained using generalised Proca theory applies, and the model is ghost free at all orders.\cite{GP} Once the degeneracy conditions (\ref{eq:degencond}) are imposed, we find that the Friedmann equations (\ref{eq:BGeqs}) become integrable and yield
\begin{equation}\label{eq:BGcorrect}
    {H} = \frac{3M_{\rm Pl}^2{H}_{\rm dS}+\eta\bar{B}_{0}^3}{3 \big(M_{\rm Pl}^2 - \xi \bar{B}_{0}^2 \big)}, \quad V(-\bar{B}_0^2) = \frac{\big(3 M_{\rm Pl}^2 {H}_{\rm dS} + \eta \bar{B}_{0}^{3}\big)^{2}}{3 \big(M_{\rm Pl}^2 - \xi \bar{B}_{0}^2 \big)},
\end{equation}
where $H_{\rm dS}$ is an integration constant and $\Lambda_B\equiv3M_{\rm Pl}^2H_{\rm dS}^2$ and so the \emph{potential is completely fixed by the background equations and is no longer arbitrary}.

For scalar perturbations, we find that after imposing the degeneracy condition (\ref{eq:degencond}), only the kinetic and gradient matrix survives and they both have rank one, meaning only one scalar degree of freedom is dynamical; the action can be written as
\begin{equation}
    S^{(2)}_S = \frac{1}{2}\int d^3x \, dt \,a^3  \Bigg[\mathcal{K}\delta\dot{B}_s^2-\mathcal{G}\frac{k^2}{a^2}\delta B_s^2\Bigg],
\end{equation}
where the kinetic coefficient vanishes identically
\begin{equation}
    \boxed{{\cal K} \equiv 0}
\end{equation}
Therefore, the scalar mode does not propagate at the linear level once the degeneracy condition (\ref{eq:degencond}) is imposed (which is necessary in order to avoid ghost modes). We stress that this result holds for any form of the potential $V_B$, and that the model is thus pathological around dynamical cosmological backgrounds.

\subsection{Stability conditions}
Once the degeneracy condition has been imposed, we can determine the exact stability conditions for the bumblebee model. First, From the observation an electromagnetic counterpart to the gravitational-wave event GW170817, it has been shown\cite{sog} that the speed of tensor modes must respect $-3\cdot 10^{-15}<c_T-1<+7\cdot 10^{-16}$. Knowing this, we assume that the error bars follow a Gaussian distribution and generate a posterior for $\xi \widetilde{\bar{B}}_0^2$ by drawing $10^4$ mock data points from this distribution, after which we find
\begin{equation}
\xi\widetilde{\bar{B}}_0^2=\left(-1.18^{+1.84}_{-1.87}\right)\cdot 10^{-15},
\end{equation}
and we can safely impose $\xi\tilde{B}_0\ll 1$. In the subhorizon limit, we have that 
$\mathcal{K}_T = 1-\xi \tilde{B}_0^2, c_T^2 = \frac{1+\xi\tilde{B}_0^2}{1-\xi\tilde{B}_0^2}$, both of which must be positive and non-zero. From this, we deduce that
\begin{equation}
    \begin{aligned}
        0 < \xi\tilde{B}_0^2 < 1 \,;& \qquad \mbox{for}\,\, \xi>0 \,,
	\\
	0 < |\xi|\tilde{B}_0^2 < 1 \,;& \qquad \mbox{for}\,\, \xi<0 \,.
    \end{aligned}
\end{equation}
Finally, we note that since $\xi\tilde{B}_0\ll 1$, the background equations can be expanded and takes the form of a quasi de Sitter solution with deviations controlled by $\xi \tilde{B}_0^2$ (small) and $\eta$, which needs not be small. In the small-$\eta$ limit, $H\approx H_{\rm dS}$ and the potential behaves as a cosmological constant at leading order.

\section{Discussion \& Conclusions}
In these proceedings, we have discussed cosmological perturbations of the most general bumblebee model on a dynamical background. By studying background evolution and perturbations, we conclude that
\begin{enumerate}
    \item the model has a higher-derivative ghost instability unless the degeneracy conditions (\ref{eq:degencond}) are imposed. Once this is done, the model is ghost-free at all orders and is a subset of generalised Proca theory with the map (\ref{eq:procamap});
    \item once the degeneracy condition is imposed, {\it the potential is no longer arbitrary} on a cosmological background;
    \item the bumblebee scalar mode does not propagate at the linear order, and the theory is therefore {\it pathological} around dynamical cosmological backgrounds.
\end{enumerate}

\section*{Acknowledgements}
N.A.N was financed by IBS under the project code IBS-R018-D3 and acknowledges support by PSL/Observatoire de Paris.

\end{document}